\documentclass[10pt,twocolumn,twoside]{IEEEtran}
\usepackage{graphicx}
\usepackage{amsfonts}
\usepackage{amssymb}
\usepackage{mathrsfs}
\usepackage{epstopdf}
\usepackage{color}
\usepackage{booktabs}
\usepackage{cite}
\usepackage{bm}
\usepackage[latin1]{inputenc}
\usepackage{amsmath}
\usepackage{amsfonts}
\usepackage{graphicx}
\usepackage{cite}
\usepackage{bm}
\usepackage{algorithmic}
\usepackage{array}
\usepackage{subfigure}
\graphicspath{{Figures/}} 
\usepackage{caption}

\begin{document}
	
	\title{Deep Learning for Signal Demodulation in Physical Layer Wireless Communications: Prototype Platform, Open Dataset, and  Analytics}
	
		\author{Hongmei Wang, Zhenzhen Wu, Shuai Ma, Songtao Lu, Han Zhang, Guoru Ding, and Shiyin Li.
\thanks{Manuscript received January  22, 2019. Corresponding
author: Shuai Ma.)}
		\thanks{H. Wang, Z. Wu, S. Ma, and S. Li are with Information and Control Engineering, China University of Mining and Technology, Xuzhou, China 221116 (Email:whm99@cumt.edu.cn, zhzhw@cumt.edu.cn, mashuai001@cumt.edu.cn, lishiyin@cumt.edu.cn).}
\thanks{S. Ma  is  also with the State Key Laboratory of Integrated Services Networks, Xidian University, Xi'an 710071, China.}
\thanks{S. Lu is with the Department of Electrical and Computer Engineering, University of Minnesota, Minneapolis, MN 55455, USA (e-mail: lus@umn.edu).}
\thanks{H. Zhang is with the Department of Electrical and Computer Engineering, University of California, Davis, CA
95616, USA (e-mail: hanzh@ucdavis.edu).}
 \thanks{G. Ding is with the College of Communications Engineering,
Army Engineering University, Nanjing 210007, China (e-mail:
 dr.guoru.ding@ieee.org)}
 	 \thanks{The work of S. Ma and S. Li were supported   by the Fundamental Research Funds for the Central Universities under Grant 2017QNA32,by
 	 	the National Natural Science Foundation  of  China under Grant 61701501, Grant 61771474; by the Natural Science Foundation of Jiangsu Province under Grant BK20170287;  by China Postdoctoral Science Foundation under Grant 2016M600452;  by
 	 	the State Key Laboratory of Integrated Services Networks (Xidian University) under grant ISN19-07; and by
 	 	the Key Laboratory of Cognitive Radio and Information Processing , Ministry of Education (Guilin University of Electronic Technology) under grant CRKL180204,   by Key Laboratory of Ocean Observation-Imaging Testbed of Zhejiang Province. The work of H. Wang was supported by the National Natural Science Foundation of China under Grant 61601464. }}

	\maketitle
	\begin{abstract}

	In this paper, we investigate   deep learning (DL)-enabled signal demodulation methods and establish the
first open dataset of real modulated signals for wireless communication systems.
Specifically, we propose a flexible communication prototype platform for  measuring real modulation dataset.
   Then, based on the measured dataset, two DL-based demodulators, called  deep belief network (DBN)-support vector machine (SVM)
   demodulator and adaptive boosting (AdaBoost) based demodulator, are proposed.
   The proposed DBN-SVM based demodulator  exploits the advantages of both DBN and SVM, i.e.,
   the advantage of DBN as a feature extractor and SVM as a feature classifier. In DBN-SVM based demodulator, the received signals are normalized before being fed to the DBN network.
       Furthermore, an AdaBoost based demodulator is developed, which employs the $k$-Nearest Neighbor (KNN)
   as a weak classifier to form a strong combined classifier. Finally, experimental results indicate that the
    proposed DBN-SVM based demodulator and AdaBoost based demodulator are superior to the single classification method using
     DBN, SVM, and maximum likelihood (MLD) based demodulator.

	\end{abstract}


	\section{Introduction}

	Conventional  wireless communication  systems are generally  designed in accordance with the rigorous    mathematical theories and  accurate  system  models \cite{John2001Digital}.
	However, because of increasing  wireless   service  requirements, such as   the use of smartphones, virtual reality, and internet of things (IoT), it is challenging to characterize future complex  wireless communication  networks
	accurately by using tractable   mathematical models or system models \cite{Shafi}.
	Recently, deep learning (DL) \cite{Lecun2015Deep},  as an effective method to handle complex  problems,
	has  attracted increasing attention from both academia and industry. DL has been  applied in image recognition \cite{Liu2016DeepFood,Zhou}, computer
	vision \cite{Geronimo}, natural language processing \cite{Tan2015Cluster} and spectrum prediction \cite{Ling2018Spectrum}, etc. In addition, some literatures have focused on using DL to optimize performance of wireless communication systems \cite{huang2019unsupervised,liu2018deep,liu2019deep}. In \cite{huang2019unsupervised}, an unsupervised learning-based fast beamforming method is proposed to maximize the weighted sum rate under the total power constraint. In \cite{liu2018deep}, a deep recurrent neural network based algorithm is proposed to tackle energy efficient resource allocation problem for heterogeneous IoT. In \cite{liu2019deep}, a three dimensional message-passing algorithm based on deep learning scheme is proposed to minimize the weighted sum of the secondary interference power for cognitive radio networks. Recent works   \cite{Timothy,Sebastian}   have interpreted  an  end-to-end  wireless communication
	system as an auto-encoder. This is promising for applications of DL to  wireless communications.

	Demodulation is one of the fundamental modules
	for wireless communications systems for high-speed transmission with a low bit error rate.
	Theoretically, optimum demodulators of  conventional wireless communication systems  are designed for
	additive white Gaussian noise (AWGN)  channels \cite{John2001Digital}.
	Moreover,  both channel state information (CSI) and channel noise distribution are usually required.
	Most previous studies \cite{Fath,Wang,Ying,Huang,Wangj,Gui} have assumed
	that each receiver can accurately estimate the fading coefficients.
	However, practical wireless communication channels
	may suffer from  multi-path fading, impulse noise, spurious
	or continuous jamming, and numerous other complex impairments,
	which deteriorate demodulation performance significantly. Because of the limited length of the training sequence, the estimate CSI will have limited accuracy \cite{Schiessl2017Delay}. Especially, for  fast-fading
	scenarios, it is difficult to accurately  estimate CSI because the fading
	coefficients   change  rapidly during the data  transmission period.
	Designing
	optimum demodulators for different channel models
	is challenging because the channel model may not
	be known at the receiver end.

	Given the above issues, DL-based model-free demodulators
	have attracted a considerable amount of attention, where the requirements for
	a \emph{priori} knowledge can be widely relaxed or even removed \cite{Mitchell2003Machine}.
	Because the information of the  modulated signals  is represented by the
	amplitude and phase, feature extraction is of critical importance for signal demodulation.
	
	DL-based demodulators have been investigated in conventional radio frequency (RF) systems.
	In \cite{mixedsignals}, a deep convolutional network demodulator
	(DCND)  is proposed  to
	demodulate mixed modulated signals, which can further reduce the bit error rate compared with the coherent demodulation method.
	In \cite{DemodulatorbasedDBN}, the authors show that the proposed   demodulator based on deep belief network (DBN) is feasible   for an AWGN channel with a certain channel impulse
	response and a Rayleigh non-frequency-selective flat fading
	channel.
	In \cite{Lanting},
	a  DL-based detector is proposed for signal demodulation in short-range  multi-channels
	without a signal equalizer.
	In \cite{fadedwireless},
	the authors show that
	deep convolutional neural networks (DCNN) for frequency-shift keying (FSK) demodulation can substantially reduce error bit probabilities  over an AWGN Rayleigh-fading channel.
	To the best of our knowledge, most of existing DL-based demodulation schemes are based on simulated data rather than real measured data.


	This paper presents a   data-driven framework for DL-based  demodulators. Specifically,
	two data-driven demodulation methods based on DBN-support vector machine (SVM) and adaptive boosting (AdaBoost) \cite{Freund1996Experiments}  are developed for end-to-end wireless communication systems. These methods learn
	and extract features    from   the received modulation signals without
	any \emph{prior} knowledge of the channel model.
	Moreover,  the performance of the  two  data-driven demodulators are evaluated on different  modulation schemes  through real measured data.
	The main contributions of this paper
	are as follows:
	
	\begin{itemize}
		
		\item A flexible end-to-end wireless communication prototype platform  is developed for application in real  physical environments,
		which  can generate real  signals.
		The prototype is used to establish measured modulation  datasets from real communication systems   in actual physical environments in eight modulation schemes, i.e., binary phase shift keying (BPSK) and multiple quadrature amplitude modulation ($M$-QAM) modulation, where $M = {2^\phi}$ and $\phi = \left\{ {2,3,4,5,6,7,8} \right\}$.  The received SNR of the eight modulated  signals are measured from
		$3$ dB to $25$ dB. An open online real modulated  dataset is established, available at
		https://pan.baidu.com/s/1biDooH6E81Toxa2u4D3p2g  or https://drive.google.com/open?id=1jXO9OMZOyVMOYv
		QSn3WVmlfQoQbonKuo
	    , where the transmission distance  of the  eight modulated  signals is measured in an indoor environment.
		To the best of our knowledge,
		this is the first open dataset of real modulated signals for wireless communication systems.

		\item Then, based on the measured data, two DL-based demodulators are proposed, namely, DBN-SVM based demodulator and AdaBoost based demodulator. The proposed DBN-SVM based demodulator, which has a novel demodulation architecture, exploits the advantages of both DBN and SVM, i.e., the advantage of DBN as a feature extractor and SVM as a feature classifier.
		To accelerate the convergence rate,
		the received signals are first normalized before being fed to the DBN network so that the features of the received signals can be extracted, the SVM is utilized to classify these features.

		\item An AdaBoost based demodulator, which utilizes multiple KNNs as a weak classifier to form a strong combined classifier, is developed.
		The proposed AdaBoost based demodulator increases the weights for
		the error demodulated symbols and decreases the corresponding weights for correctly demodulated
		symbols during the iterations.

		\item Finally, the demodulation performance of the two proposed data-driven demodulators are investigated.
		Specifically, the demodulation accuracies of the two DL-based demodulators decrease over the respective transmission and modulation orders for a fixed transmission distance.
		The experimental results also show that the demodulation accuracy of the
		DBN-SVM based demodulator is higher than those of DBN-based and SVM-based demodulators.
		Moreover, the demodulation accuracy of the
		AdaBoost based demodulator is higher than that of the DBN-SVM based demodulator at the lower SNR regions, and the accuracies of the two demodulators are similar at high SNRs.
		For the high SNR scenario,
		a high-order modulation is generally preferred.

	\end{itemize}
	
	The remainder of this paper is organized as follows. Section II describes the system model. Section III explores the structures of the DBN-SVM and AdaBoost, including detailed descriptions of the data stream and how to make classification decisions. In section IV, the data analysis results are provided and analyzed. Finally, the conclusions from the study are drawn in Section V.

	\emph{Notations}: Boldfaced lowercase and uppercase letters represent vectors and matrices, respectively. The transpose of a matrix is denoted as ${\left(  \cdot  \right)^{\rm{T}}}$. $\mathcal{L}\buildrel \Delta \over =\{1, 2, ...,L\}$, $ \mathcal{L}_1\buildrel \Delta \over=\{1,...,L_1\}$,  $\mathcal{M}_k \buildrel \Delta \over = \left\{ {1,2,...,{M_k}} \right\}$, $\mathcal{N}_k \buildrel \Delta \over = \left\{ {1,2,...,{N_k}} \right\}$, $\mathcal{D}\buildrel \Delta \over =\{1,...,D\}$, $\mathcal{Q}\buildrel \Delta \over=\{0,1, \ldots ,\bar M\} $, and  $\mathcal{M}\buildrel \Delta \over =\left\{ {0,1, \ldots ,M - 1} \right\}$.

	\section{System Model}
	
\begin{figure}[h]
		\centering
		\includegraphics[width=0.45\textwidth]{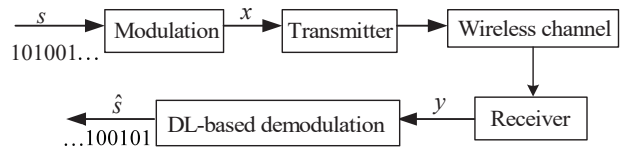}
		\caption{End-to-end wireless system model}
		\label{fig 1}
	\end{figure}

	An end-to-end wireless communication system\footnote{The term end-to-end   wireless system  model implies that
		signal features are learned from a single deep neural network,
		without the  complex multi-stage expert machine learning processing\cite{Timothy,Sebastian,end-to-end,Karanov,Lecun2006Off}.} is considered, which includes a single antenna transmitter and a single antenna receiver, as illustrated in Fig. \ref{fig 1}. By adopting the BPSK or $M$-QAM digital modulation schemes,
	the transmitted signal $x\left( t \right)$ is given as
	\begin{align}
	x\left( t \right) = {V_m}\cos \left( {2\pi {f_c}t + {\theta _m}} \right),\quad m = 1,...,M,\quad 1 \le t \le T,
	\end{align}
	where ${V_m}$, ${{\theta _m}}$ and $T$ denote
	the amplitude, phase, and period of the signal $x\left( t \right)$, respectively;
	${f_c}$ is the carrier frequency. 

	Let $g\left( t \right)$ denote the multipath channel between the transmitter and the receiver, which may
	suffer nonlinear distortion, interference, and frequency selective fading.
	At the receiver, the received signal $y\left( t \right)$ is given by
	\begin{align}
	{y\left( t \right) = g\left( t \right)x\left( t \right) + {n_r}\left( t \right)},
	\end{align}
	where ${n_r}\left( t \right)$ denotes the received noise.
	
	Then, the received analog signal $y\left( t \right)$ is converted to the digital signal via the vector signal analyzer. Let ${\bf{y}} \buildrel \Delta \over = {\left[ {{y_1},{y_2},...,{y_{NL}}} \right]^T}$ denote the total sampled digital signal vector, where
	${y_n} = y\left( {\frac{{n - 1}}{N}T} \right)$ is the $n$th sample, $N$ is the  number of samples of one period, and $L$ denotes the number of signal periods.

	Before the demodulation process, the received signal ${\bf{y}}$ is normalized to $[0,1]$, which can accelerate the DL network processing speed\cite{Importance}.
	Senerally, the normalized data ${\bf{\hat y}} \buildrel \Delta \over = {\left[ {{{\hat y}_1},{{\hat y}_2},...,{{\hat y}_{NL}}} \right]^T}$
	is given by
	\begin{align}
	{{\hat y_i} = \frac{{{y_i} - {y_{\min }}}}{{{y_{\max }} - {y_{\min }}}},\quad 1 \le i \le N{L}},
	\end{align}
	where ${y_{\min }} = \mathop {\min }\limits_{1 \le i \le N{L}} {y_i}$, and ${y_{\max }} = \mathop {\max }\limits_{1 \le i \le N{L}} {y_i}$.

	Because the information of the BPSK and $M$-QAM are represented by amplitudes and phases,
	DL is used to extract information features from the received signals.
	Specifically, with the sampled signal vector ${\mathbf{y}}$, two DL-based demodulators are proposed: DBN-SVM based demodulator and AdaBoost based demodulator.
	The DL-based demodulators consist of two phases: training phase and testing phase.
	During the training phase, the parameters of the DL-based demodulators are optimized with the training dataset.
	Then, in the testing phase, the demodulators demodulate the received signal and recover the transmitted information.
	
	Let ${{z_l}}$ denote the label signal of the $l$th period, where $l \in \mathcal{L}$ and $\Phi $ is the label set,
	i.e., $\Phi = \left\{ {{z_1},{z_2}, \ldots ,{z_L}} \right\}$, which is determined by the modulation scheme. 
	Let ${\rm{{\cal T}_1}} = \left\{ {\left( {{{{\bf{\hat y}}}_1},{z_1}} \right),\left( {{{{\bf{\hat y}}}_2},{z_2}} \right), \ldots ,\left( {{{{\bf{\hat y}}}_{{L_1}}},{z_{{L_1}}}} \right)} \right\}$ denote the labeled training signal set,
	where ${{{\bf{\hat y}}}_l} = {\left[ {{{\hat y}_{1 + \left( {l - 1} \right)N}},{{\hat y}_{2 + \left( {l - 1} \right)N}},...,{{\hat y}_{lN}}} \right]^T}$
	denotes the normalized signal of the $l$th period, and $L_1$ denotes the total number of training signal periods $\left( {{L_1} < L} \right)$.

	\section{DBN-SVM based Demodulator}

\begin{figure}
	\centering
	\includegraphics[width=0.32\textwidth]{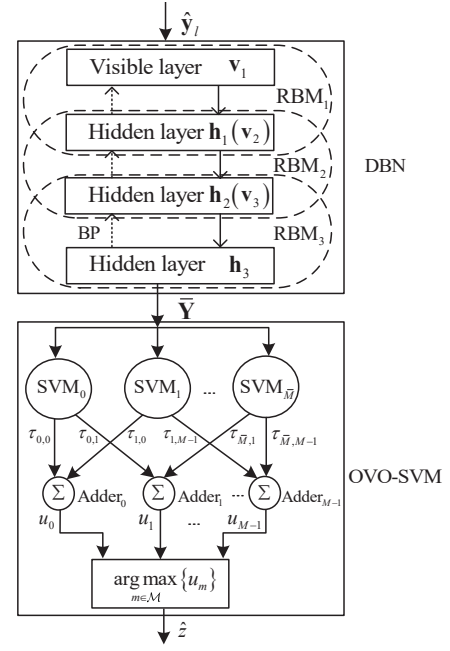}
	\caption{Structure of DBN-SVM based demodulator}
	\label{DBN_DAGSVM}
\end{figure}

	As
	an unsupervised features extraction method, the DBN can efficiently extract high-level and hierarchical
	features from the measured signal, while the SVM minimizes the structure
	risk and shows good learning and generalization performance with a small amount of samples.
	Inspired by those advantages of the two approaches, a combination of DBN and SVM for demodulation is proposed. The DBN-SVM demodulator is shown in Fig. \ref{DBN_DAGSVM}, the DBN is used as a feature generator and the SVM is used as a classifier.
	
	\subsubsection{DBN}
	
	The proposed DBN includes three stacked restricted Boltzmann machines (RBM)\cite{Hinton2002Training}, i.e., RBM$_1$, RBM$_2$, and RBM$_3$,
	as shown in Fig. \ref{DBN_DAGSVM}.
	Specifically, RBM$_k$ is an undirected,
	bipartite graphical model, and it composes a visible
	layer ${{\bf{v}}_k} = [v_{k,1},v_{k,2},...,v_{k,{M_k}}]^T$ and a hidden layer ${{\bf{h}}_k} = [h_{k,1},h_{k,2},...,h_{k,{N_k}}]^T$, where $v_{k,\alpha}$ and $h_{k,\beta}$ are the $\alpha$th neuron of ${{\bf{v}}_k}$ and the $\beta$th neuron of ${{\bf{h}}_k}$, respectively, $\alpha \in \mathcal{M}_k$, $\beta \in \mathcal{N}_k$, $k \in \left\{ {1,2,3} \right\}$. The visible layer ${{\bf{v}}_k}$ and hidden layer ${\bf{h}}_k$ are fully connected via a symmetric undirected weighted matrix ${{\bf{W}}_k} = \left[ {{{\bf{w}}_{k,1}},{{\bf{w}}_{k,2}}, \ldots ,{{\bf{w}}_{k,{N_k}}}} \right]^T$, where ${{\bf{w}}_{k,\beta}} = {[{w_{k,\beta}^{\left( 1 \right)},w_{k,\beta}^{\left(2 \right)}, \ldots ,w_{{k},\beta}^{\left( M_k \right)}}]^T}$ is a weight vector between ${{\bf{v}}_k}$ and ${h_{k,\beta}}$.
	For the three RBM, there is no intralayer connections between either
	the visible layer or the hidden layer.
	
	For RBM$_k$, the energy $E\left( {{{\bf{v}}_k},{{\bf{h}}_k}} \right)$ is defined by combining the configuration of both  ${{\bf{v}}_k}$ and
	${{\bf{h}}_k}$ as follows
	\begin{align}
	E\left( {{{\bf{v}}_k},{{\bf{h}}_k}} \right) =  - {\bf{h}}_k^T{{\bf{W}}_k}{{\bf{v}}_k} - {\bf{a}}_k^T{{\bf{v}}_k} - {\bf{b}}_k^T{{\bf{h}}_k},
	\end{align}
	where ${{\bf{a}}_k} = {\left[ {a_{k,1},a_{k,2},...,a_{k,{{M_K}}}} \right]^T}$ is an offset vector of ${{\bf{v}}_k}$,
	and ${{\bf{b}}_k} = {\left[ {b_{k,1},b_{k,2},...,b_{k,{{N_K}}}} \right]^T}$ is an offset vector
	of ${{\bf{h}}_k}$.
	
	Based on $E\left( {{{\bf{v}}_k},{{\bf{h}}_k}} \right)$, the probability of  ${{\bf{v}}_k}$ is given by
	\begin{align}
	p\left( {{{\bf{v}}_k}} \right) = \frac{1}{Z_k}\sum\limits_{{{\bf{h}}_k}} {{e^{E\left( {{{\bf{v}}_k},{{\bf{h}}_k}} \right)}}} ,
	\end{align}
	where ${Z_k} = \sum\limits_{{{\bf{v}}_k},{{\bf{h}}_k}} {{e^{E\left( {{{\bf{v}}_k},{{\bf{h}}_k}} \right)}}} $ is the normalization factor.
	
	During the training phrase, the goal of the RBM$_k$ is to maximize
	the log-likelihood function as follows
	\begin{align}\label{goal}
	\mathop {\max }\limits_{{{\bf{W}}_k},{{\bf{a}}_k},{{\bf{b}}_k}} \sum\limits_{{{\bf{v}}_k}} {\log p\left( {{{\bf{v}}_k}} \right)} .
	\end{align}
	
	To solve equation \eqref{goal}, the gradient descent method is used to iteratively calculate the variables ${\bf{W}}_k$, ${\bf{a}}_k$, and ${\bf{b}}_k$, where the corresponding partial derivative with respect to $\mathbf{W}_k$, $\mathbf{a}_k$, and $\mathbf{b}_k$ can be written as
	
	\begin{subequations}
		\begin{align}
		\frac{{\partial \log p\left( {{{\bf{v}}_k}} \right)}}{{\partial w_{k,\beta}^{\left( \alpha \right)}}} &= v_{k,\alpha}p\left( {h_{k,\beta} = 1|{{\bf{v}}_k}} \right)\nonumber \\
		- \sum\limits_{{{\bf{v}}_k}} p\left( {{{\bf{v}}_k}} \right)&p\left( {h_{k,\beta} = 1|{{\bf{v}}_k}} \right)v_{k,\alpha}, \quad \alpha \in \mathcal{M}_k, \beta \in \mathcal{N}_k,  \\
		\frac{{\partial \log p\left( {{{\bf{v}}_k}} \right)}}{{\partial a_{k,\alpha}}} &= v_{k,\alpha} -\sum\limits_{{{\bf{v}}_k}} {p\left( {{{\bf{v}}_k}} \right)} v_{k,\alpha},\quad \alpha \in \mathcal{M}_k,\\
		\frac{{\partial \log p\left( {{{\bf{v}}_k}} \right)}}{{\partial b_{k,\beta}}} &= p\left( {h_{k,\beta} = 1|{{\bf{v}}_k}} \right) \nonumber\\
		- \sum\limits_{{{\bf{v}}_k}} p\left( {{{\bf{v}}_k}} \right)&p\left( {h_{k,\beta} = 1|{{\bf{v}}_k}} \right), \quad \beta \in \mathcal{N}_k.
		\end{align}
	\end{subequations}
	
	According to \cite{DAEandDBN}, the conditional probability $p\left( {h_{k,\beta} = 1|{{\bf{v}}_k}} \right)$ and $p\left( v_{k,\alpha} = 1\left| {{{\bf{h}}_k}} \right. \right)$ are respectively given by
	\begin{subequations}
		\begin{align}
		&p\left( {h_{k,\beta} = 1|{{\bf{v}}_k}} \right) = {\rm{sigmoid}}\left( {b_{k,\beta} + {{\bf{v}}_k}^T{\bf{w}}_{k,\beta}} \right), \\
		&p\left( {v_{k,\alpha} = 1\left| {{{\bf{h}}_k}} \right.} \right) = {\rm{sigmoid}}\left( {a_{k,\alpha} + \sum_{\mathit{\beta}=1}^{N_k} {{h_{k,\beta}}{{w}_{k,\beta}^{(\alpha)}}}} \right),
		\end{align}
	\end{subequations}
	where ${\rm{sigmoid}}\left( x \right) \buildrel \Delta \over = \frac{1}{{1 + {e^{ - x}}}}$, $\alpha \in \mathcal{M}_k$, $\beta \in \mathcal{N}_k$, $h_{k,\beta}$, and $v_{k,\alpha}  \in \left[ {0,1} \right]$.
	
	Then, the variables ${\bf{W}}_k$, ${\bf{a}}_k$, and ${\bf{b}}_k$ are updated by the following equations\cite{Hinton1}
	\begin{subequations}
		\begin{align}
		w_{k+1,\beta}^{\left( \alpha \right)}&\leftarrow w_{k,\beta}^{\left( \alpha \right)} + \eta \frac{{\partial \log p\left( {{{\bf{v}}_k}} \right)}}{{\partial w_{k,\beta}^{\left( \alpha \right)}}}, \quad \alpha \in \mathcal{M}_k, \beta \in \mathcal{N}_k, \\
		a_{k+1,\alpha}& \leftarrow a_{k,\alpha} + \eta \frac{{\partial \log p\left( {{{\bf{v}}_k}} \right)}}{{\partial a_{k,\alpha}}}, \quad \alpha \in \mathcal{M}_k,\\
		b_{k+1,\beta} &\leftarrow b_{k,\beta} + \eta \frac{{\partial \log p\left( {{{\bf{v}}_k}} \right)}}{{\partial b_{k,\beta}}}, \quad \beta \in \mathcal{N}_k,
		\end{align}
	\end{subequations}
	where $\eta > 0$ is the learning rate.
	
	By employing the gradient descent method, RBM$_1$ is trained first, where ${{\bf{v}}_1} = {\bf{\hat y}}_l$ and $l \in \mathcal{L}_1$. Then, let ${{\bf{v}}_2}={{\bf{h}}_1}$, and RBM$_2$ is trained. Similarly, after training RBM$_2$, let ${{\bf{v}}_3}={{\bf{h}}_2}$, and RBM$_3$ is trained. Moreover, when RBM$_3$ is trained, the parameters of DBN can be obtained, i.e., ${\left\{ {{{\bf{W}}_k},{{\bf{a}}_k},{{\bf{b}}_k}} \right\}_{k \in\{1,2,3\}}}$. Then, the parameters ${\left\{ {{{\bf{W}}_k},{{\bf{a}}_k},{{\bf{b}}_k}} \right\}_{k \in\{1,2,3\} }}$ are further fine-tuned by the supervised back propagation (BP) algorithm \cite{Zhang}.
	
	After DBN is trained, it outputs the extracted feature ${{{\bf{\bar y}}}_{l_1}}= {{\bf{h}}_3}$, where $l_1 \in \mathcal{L}_1$.
	Let ${\bf{\bar Y}} = {\left[ {{{{\bf{\bar y}}}_1},{{{\bf{\bar y}}}_2}, \ldots ,{{{\bf{\bar y}}}_{{L_1}}}} \right]^T}$ denote the output feature set.

	\subsubsection{OVO-SVM}

	With the extracted feature set ${\bf{\bar Y}} $, the one-versus-one (OVO)-SVM is adopted for further classification, which achieves multiclassification by solving the two-classification subproblems \cite{Wang2009Multi,Marginclassif1999}.
	As shown in Fig. \ref{DBN_DAGSVM},
	OVO-SVM exploits ${\bar M}$ nonlinear two-class SVMs, i.e., SVM$_0$,...,SVM$_{\bar M}$, to classify
	$M$ categories for $M$-QAM modulation, where $\bar M \buildrel \Delta \over = \frac{{M\left( {M - 1} \right)}}{2} - 1$.

	To map pedestrian features to a high dimensional space,  a Gaussian kernel is introduced, which can be expressed as
	\begin{align}
	{G_q}\left( {{{{\bf{\bar y}}}_{l_1}},{{{\bf{\bar y}}}_{l_2}}} \right) = \exp \left( { - \frac{{{{\left\| {{{{\bf{\bar y}}}_{l_1}} - {{{\bf{\bar y}}}_{l_2}}} \right\|}^2}}}{{2\sigma _q^2}}} \right),
	\end{align}
	where $\sigma_q > 0$ is the bandwidth of the Gaussian kernel and $q \in\mathcal{Q}$.

	According to the nonlinear SVM theory \cite{Rhuma2011An}, the nonlinear two-class SVM$_q$ problem can be formulated as
	\begin{subequations}\label{linear_p}
		\begin{align}
		\mathop {\min }\limits_{{{\bf{c}}_q}} {\rm{ }}\quad & \frac{1}{2}\sum\limits_{l_1 = 1}^{{L_1}} {\sum\limits_{l_2 = 1}^{{L_1}} {{c_{q,l_1}}{c_{q,l_2}}{z_{l_1}}{z_{l_2}}{G_q}\left( {{{{\bf{\bar y}}}_{l_1}},{{{\bf{\bar y}}}_{l_2}}} \right)}  - \sum\limits_{{l_1} = 1}^{{L_1}} {{c_{q,l_1}}} }       \\
		{\rm{s}}.{\rm{t}}. {\rm{ }}\quad &\sum\limits_{l_1 = 1}^{{L_1}} {{c_{q,l_1}}{z_{l_1}} = 0},\\
		&0 \le {c_{q,{l_1}}} \le K,\quad l_1 \in \mathcal{L}_1,
		\end{align}
	\end{subequations}
	where ${{\bf{c}}_q} = {\left[ {{c_{q,1}},{c_{q,2}}, \ldots ,{c_{q,{L_1}}}} \right]^T}$ and $q \in\mathcal{Q}$.
	
	By solving linear programming \eqref{linear_p}, the optimal solution ${\bf{c}}_q^ * = {\left[ {c_{q,1}^*,c_{q,2}^*, \ldots ,c_{q,{L_1}}^*} \right]^T}$ is obtained.
	Then, the nonlinear two-class SVM$_q$ decision function ${f_q}\left( {{{{\bf{\bar y}}}_{l_1}}} \right)$, with $l_1 \in \mathcal{L}_1$, $q \in\mathcal{Q}$, is given as
	\begin{align}\label{dec}
	{f_q}\left( {{{{\bf{\bar y}}}_{l_1}}} \right) = \gamma \left( {\sum\limits_{i = 1}^{{L_1}} {c_{q,i}^*} {z_i}\exp \left( { - \frac{{{{\left\| {{{{\bf{\bar y}}}_i} - {{{\bf{\bar y}}}_{l_1}}} \right\|}^2}}}{{2\sigma _q^2}}} \right) + b_q^*} \right),
	\end{align}
	where $b_q^* \buildrel \Delta \over = {z_{l_1}} - \sum\limits_{i = 1}^{{L_1}} {c_{q,{l_1}}^*} {z_i}\exp \left( { - \frac{{{{\left\| {{{{\bf{\bar y}}}_i} - {{{\bf{\bar y}}}_{l_1}}} \right\|}^2}}}{{2\sigma _q^2}}} \right)$ is a biased variable\cite{Wang2018Research},
	and
	\begin{align}\gamma \left( x \right) \buildrel \Delta \over = \left\{ {\begin{array}{*{20}{l}}
		{1,\quad \rm{if}\quad x \ge 0}\\
		{0,\quad \rm{if}\quad x < 0}
		\end{array}} \right..
	\end{align}
	
	Let $\tau _{q,\bar m}$ and $\tau _{q,m}$ denote the output of the SVM$_q$, and $\tau_{q,\bar m}$ and $\tau_{q,m}$ are the inputs of the Adder$_{\bar m}$ and the Adder$_m$, respectively,
	where ${\tau _{q,\bar m}},{\tau _{q,m}} \in \left\{ {0,1} \right\}$, ${\tau _{q,\bar m}} + {\tau _{q,m}} = 1$, ${\bar m}$, $m \in \mathcal{M}$, and ${\bar m} \ne m$. Then, for Adder${_{ m}}$, the number of votes is updated by ${u_m} = {u_m} + {\tau _{q,m}}$, where
	the initial value of ${u_m}$ is $0$, and $m \in \mathcal{M}$.

	Then, with the number of votes $\left\{ {{u_m}} \right\}_{m = 0}^{M - 1}$, the output label $\hat z$ is obtained as follows
	\begin{align}
	\hat z = \mathop {\arg \max }\limits_{m \in \mathcal{M}} \left\{ {{u_m}} \right\}.
	\end{align}
	Finally, $\hat z$ is mapped to the demodulation result $\hat s$.
	
	After the entire network is trained, the parameters ${\bf{W}}_k$ , ${\bf{a}}_k$, and ${\bf{b}}_k$ of the DBN, and ${\bf{c}}_q,b_q^*,\sigma_q $ of the OVO-SVM are optimized, where $k\in\{1,2,3\}$ and $q \in\mathcal{Q}$.
	Then, the test signal ${\rm{{\cal T}_2}} = \left\{ {\left( {{{{\bf{\hat y}}}_{L_1+1}},{z_{L_1+1}}} \right),\left( {{{{\bf{\hat y}}}_{L_1+2}},{z_{L_1+2}}} \right), \ldots ,\left( {{{{\bf{\hat y}}}_{{L}}},{z_{{L}}}} \right)} \right\}$ is converted to the feature vector ${{\bf{\bar y}}}$, where ${L_2}$ is the number of test signal periods.
	The details of the DBN-SVM based demodulator are listed in Algorithm $1$.
		\begin{table}
		\begin{center}
			\hrule
			\vspace{0.3cm}
			\begin{enumerate}
				\textbf{Algorithm 1}: The DBN-SVM based demodulator \\
				\vspace{0.2cm}
				\hrule
				\vspace{0.2cm}
				1.\quad Given the labeled training signal ${\bf{\hat y}}_l$, $l \in \mathcal{L}_1$;\\
				2.\quad Initialize ${\bf v}_1$, ${\bf h}_1$, ${{\bf W}_1}$, ${\bf a}_1$ and ${\bf b}_1$;\\
				3.\quad For $k=1,\ldots,3$ do\\
				4.\quad \quad Train $k$th RBM according to formula $\left( 4 \right) - \left( 6 \right)$;\\
				5.\quad \quad Update $ {{\bf{W}}_k},{{\bf{a}}_k}$ and ${{\bf{b}}_k}$ are according to formula $\left( 7 \right) - \left( 9 \right)$;  \\
				6.\quad End for;\\
				7.\quad  Get the extracted feature vector set ${\bf{\bar Y}}$, and classified by \\
				\quad \quad \quad OVO-SVM;\\
				8.\quad Update Adder$_m$, $m \in \mathcal{M}$;\\
				9.\quad Output label $\hat z$: \\
				\quad \quad\quad \quad \quad $\hat z = \mathop {\arg \max }\limits_{m \in \mathcal{M}} \left\{ {{u_m}} \right\}$.
			\end{enumerate}
			\vspace{0.2cm}
			\hrule
			\label{algorithm1}\end{center}
	\end{table}

	\section{AdaBoost Based Demodulator}

	AdaBoost is a general method used to improve machine learning algorithms \cite{Mcdonald2003An}, which integrates multiple independent weakly classifiers into a stronger classifier. In this section, we exploit the $k$-Nearest Neighbor (KNN) classifiers as the weak classifier for constructing the AdaBoost.
	
	\begin{figure}[h]
		\centering
		\includegraphics[width=0.3\textwidth]{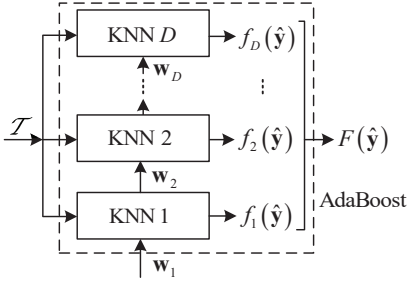}
		\caption{Structure of AdaBoost based demodulator }
		\label{adaboost}
	\end{figure}
	As shown in Fig. \ref{adaboost}, the proposed AdaBoost consists $D$ KNN classifiers.
	The labeled training signal set is denoted by ${\rm{{\cal T}}} = \left\{ {\left( {{{{\bf{\hat y}}}_1},{z_1}} \right),\left( {{{{\bf{\hat y}}}_2},{z_2}} \right), \ldots ,\left( {{{{\bf{\hat y}}}_{{L_1}}},{z_{{L_1}}}} \right)} \right\}$.

	Let ${{\bf{w}}_d} = {\left[ {{w_d}\left( 1 \right),{w_d}\left( 2 \right), \ldots ,{w_d}\left( L_1 \right)} \right]^T}$ denote the weight vector of $d$th KNN, where $0 \le {w_d}\left( l \right) \le 1$, $l \in \mathcal{L}_1$, and $\sum\limits_{l = 1}^{{L_1}} {{w_d}\left( l \right)} = 1$, $d \in \mathcal{D}$. For the $1$st KNN, ${w_1}\left( l \right) = \frac{1}{L_1}$, $l \in \mathcal{L}_1$.
	Based on the weight vector ${\bf{w}}_d$, the $d$th KNN re-samples the training set ${\rm{{\cal T}}}$ and generates a new training
	set ${{{\rm{{\cal T}}}_{d}}} = \left\{ {\left( {{{{\bf{\tilde y}}}_{d_1}},{z_{d_1}}} \right),\left( {{{{\bf{\tilde y}}}_{d_2}},{z_{d_2}}} \right), \ldots ,\left( {{{{\bf{\tilde y}}}_{{d_{L1}}}},{z_{{d_{L1}}}}} \right)} \right\}$, $ d_{l} \in \mathcal{L}_1$.
	
	Then, a vector in ${{{\rm{{\cal T}}}_{d}}}$ is searched with the minimum distance from ${{{{\bf{\hat y}}}_{l}}}$, i.e.,
	\begin{align}
	l^* = \mathop {\arg \min }\limits_{i \in {\mathcal{L}_1}} {\left\| {{{{\mathbf{\tilde y}}}_{d_i}} - {{{\mathbf{\hat y}}}_{{l}}}} \right\|_2},\quad l \in \mathcal{L}_1,d\in \mathcal{D}.
	\end{align}
	
	
	
	Because the label of ${{\bf{\tilde y}}_{{d_l^ * }}}$ is ${ z_{d_l^ *}}$, ${f_d}\left( {{{\bf{\hat y}}_{l }}} \right) = {{ z_{d_l^ *}}}$, which implies that the classification result of $d$th KNN  for ${{{{\bf{\hat y}}}_{l }}}$ is ${ z_{d_l^ *}}$.
	
	Let ${\chi _d}$ denote the weight sum of misclassified samples of $d$th KNN as follows
	\begin{align}
	{\chi _d} = \sum\limits_{{l} = 1}^{L_1} {{w_d}\left( {l } \right)I\left( {{f_d}\left( {{{\bf{\hat y}}_{l }}} \right),{ z_{l }}} \right)},\quad d\in \mathcal{D},
	\end{align}
	where $I\left( {x,y} \right) $ is the indicator function, i.e.,
	\begin{align}
	I\left( {x,y} \right) = \left\{ {\begin{array}{*{20}{l}}
		{1,\quad {\text{if}} \quad x \ne y}\\
		{0,\quad {\text{if}} \quad x = y}.
		\end{array}} \right.
	\end{align}

	Then,
	for $(d+1)$th KNN, weight ${{\bf{w}}_{(d+1)}} = {\left[ {{w_{(d+1)}}\left( 1 \right), \ldots ,{w_{(d+1)}}\left( L_1 \right)} \right]^T}$ is updated as
	\begin{align}
	{w_{d + 1}}\left( {l } \right) = \frac{{{w_d}\left( {l} \right)\exp \left( { - {\alpha _d}I\left( {{f_d}\left( {{{{\bf{\hat y}}}_{l}}} \right),{ z_{l}}} \right)} \right)}}{{{Q_d}}},\notag\\
	\qquad \qquad\qquad\qquad \qquad l \in \mathcal{L}_1,d\in \mathcal{D},
	\end{align}
	where ${\alpha _d} = \frac{1}{2}\ln \left( {\frac{{1 - {\chi _d}}}{{{\chi _d}}}} \right)$, and ${Q_d} = \sum\limits_{{l} = 1}^{{L_1}} {{w_d}\left( {l} \right)\exp \left( { - {\alpha _d}I\left( {{f_d}\left( {{{{\bf{\hat y}}}_{l}}} \right),{z_{l }}} \right)} \right)} $ is the normalization factor. If $ {{{\bf{\hat y}}}_{l}}$ is classified correctly, i.e., $I\left( {{f_d}\left( {{{{\bf{\hat y}}}_{l}}} \right),{z_{l}}} \right) = 0$, ${w_{d + 1}}\left( {l} \right) = \frac{{{w_d}\left( {l} \right)}}{{{Q_d}}}$.
	Otherwise, ${w_{d + 1}}\left( {l} \right) = \frac{{{w_d}\left({l} \right)\exp \left( { - {\alpha _d}} \right)}}{{{Q_d}}}$.
	
	
	After training $D$ KNNs, AdaBoost classifies ${{{{\bf{\hat y}}}_{l}}}$ as follows
	\begin{align}
	F\left( {{{{\bf{\hat y}}}_{l}}} \right) = \hat {{z_l}} = \mathop {\arg \max }\limits_{z_l \in \Phi } {\sum\limits_{d = 1}^D {{\alpha _d}\left( {1 - I\left( {{f_d}\left( {{{{\bf{\hat y}}}_{l}}} \right),z_l} \right)} \right)} } ,
	\end{align}
	where ${{\alpha _d}}$ is the coefficient of $\left( {1 - I\left( {{f_d}\left( {{{{\bf{\hat y}}}_{l}}} \right),z_l} \right)} \right)$ and $I\left( {{f_d}\left( {{{{\bf{\hat y}}}_{l}}} \right),z_l} \right)$ can be regarded as the voting value, i.e., if $I\left( {{f_d}\left( {{{{\bf{\hat y}}}_{l}}} \right),z_l} \right)=0$, ${{f_d}\left( {{{{\bf{\hat y}}}}} \right)}$ classifies signal ${{{{\bf{\hat y}}}_{l}}}$ into class $z_l$, otherwise, ${{{{\bf{\hat y}}}_{l}}}$ does not belong to class $z_l$. The class with the maximum sum of weighted voting value, ${{\alpha _d}\left( {1 - I\left( {{f_d}\left( {{{{\bf{\hat y}}}_{l}}} \right),z_l} \right)} \right)} $, for all classifiers, is identified as the classification result ${\hat z_{{l}}}$ of the Adaboost classifier, and then ${\hat z_{{l}}}$ is mapped to demodulation result ${\hat s_{{l}}}$.
	The details of the KNN-based AdaBoost demodulator are listed in Algorithm $2$.
	
	\begin{table}
		\begin{center}
			\hrule
			\vspace{0.3cm}
			\begin{enumerate}
				\textbf{Algorithm 2}: The KNN based AdaBoost demodulator \\
				\vspace{0.2cm}
				\hrule
				\vspace{0.2cm}
				1.\quad Given the labeled training signal set ${\rm{{\cal T}}}$ ;\\
				2.\quad Initialize signal weight ${w_1}\left( {l} \right) = \frac{1}{L_1}$;\\
				3.\quad For $d=1,\ldots,D$ do\\
				4.\quad \quad Train $d$th KNN according to weights ${{\bf{w}}_d}$;\\
				5.\quad \quad Get weak classifier ${f_d}\left( {{{{\bf\hat y}}_{l}}} \right) \in \mathcal{M}$ with error rate \\
				\quad \quad \quad \quad ${\chi _d} = \sum\limits_{{l} = 1}^{L_1} {{w_d}\left({l} \right)I\left( {{f_d}\left( {{{{\bf \hat y}}_{l}}} \right),{z_{l}}} \right)}$\quad $d \in \mathcal D$;\\
				6.\quad \quad Update:\\
				\quad \quad \quad \quad ${w_{d + 1}}\left( {l} \right) = \frac{{{w_d}\left({l} \right)\exp \left( { - {\alpha _d}I\left( {{f_d}\left( {{{{\bf{\hat y}}}_{l}}} \right),{ z_{l}}} \right)} \right)}}{{{Q_d}}},\quad l \in \mathcal{L}_1$\\
				\quad \quad \quad\quad where ${\alpha _d} = \frac{1}{2}\ln \left( {\frac{{1 - {\chi _d}}}{{{\chi _d}}}} \right)$, and \\ \quad \quad \quad\quad${Q_d} = \sum\limits_{{l} = 1}^{{L_1}} {{w_d}\left( {l} \right)\exp \left( { - {\alpha _d}I\left( {{f_d}\left( {{{{\bf{\hat y}}}_{l}}} \right),{z_{l}}} \right)} \right)}$;\\
				7.\quad End for;\\
				8.\quad Output the final decision classifier:\\
				\quad \quad \quad \quad $F\left( {{{{\bf{\hat y}}}_{l}}} \right) = \hat {{z_l}} = \mathop {\arg \max }\limits_{z \in \Phi } {\sum\limits_{d = 1}^D {{\alpha _d}\left( {1 - I\left( {{f_d}\left( {{{{\bf{\hat y}}}_{l}}} \right),z_l} \right)} \right)} }$.
			\end{enumerate}
			\vspace{0.2cm}
			\hrule
			\label{algorithm1}\end{center}
	\end{table}

	\section{Experimental Results and Discussions}
	In this section, the performance of the proposed DBN-SVM based demodulator and AdaBoost based demodulator is investigated. Also the performance of the the DBN based, SVM based, and maximum likelihood (MLD) based demodulation methods are presented for comparison.
	
	\subsection{The end-to-end wireless communication system prototype}

\begin{figure}[tbp]
	\begin{minipage}[t]{0.5\linewidth}
		\centering
		\includegraphics[height=6cm,width=4.1cm]{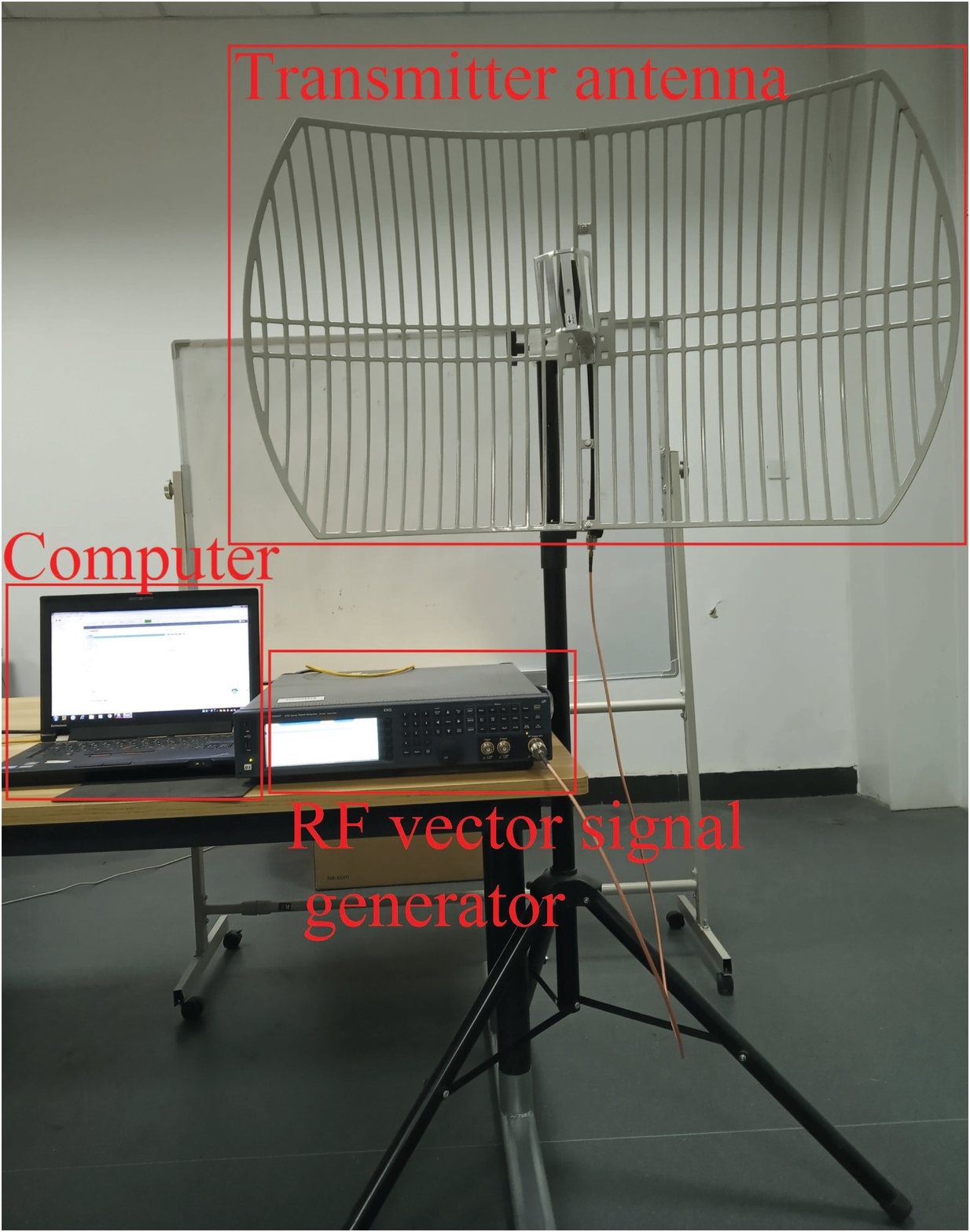}
	\end{minipage}%
	\begin{minipage}[t]{0.5\linewidth}
		\centering
		\includegraphics[height=6cm,width=3.8cm]{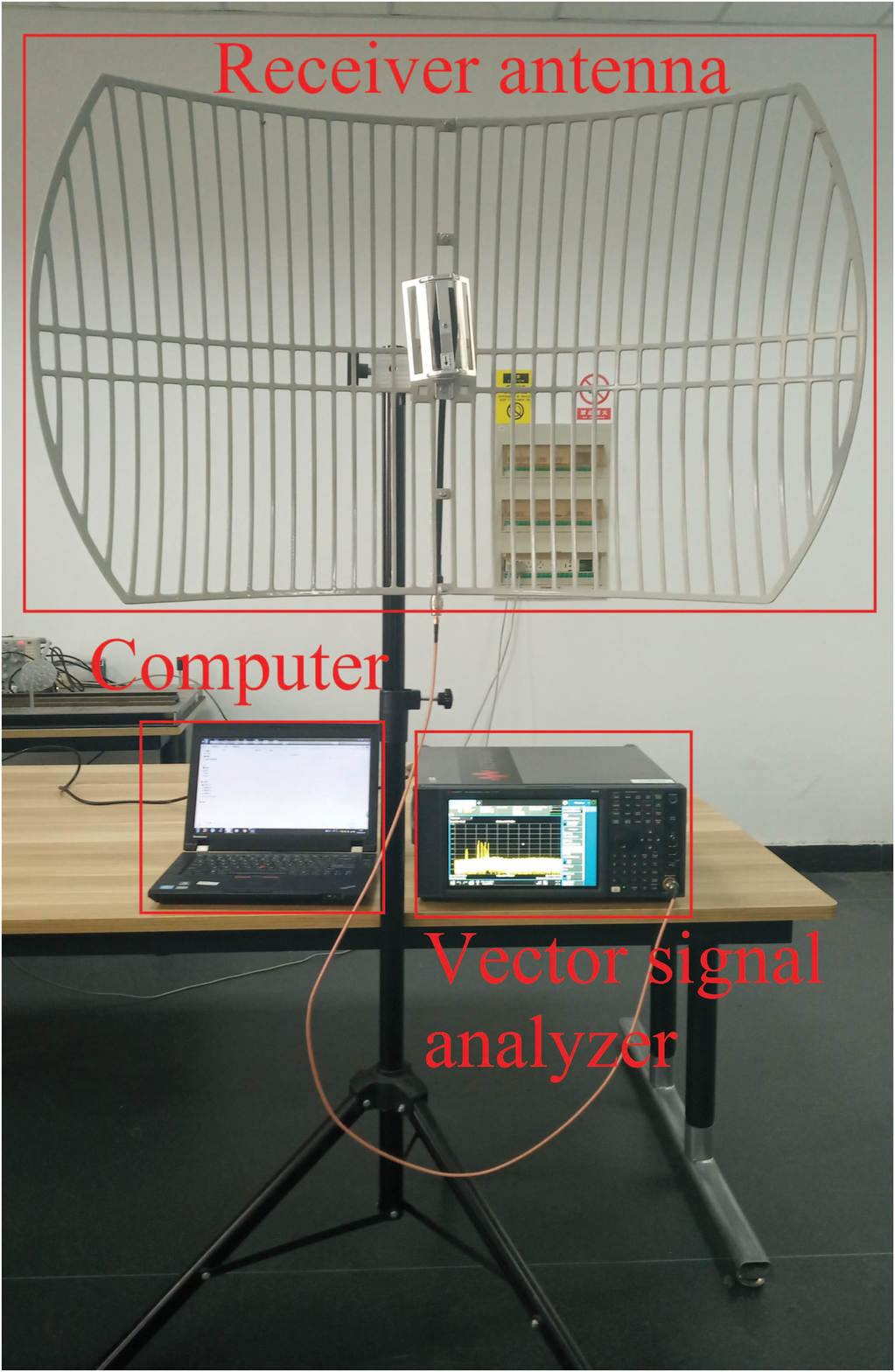}
	\end{minipage}
	\caption{End-to-end wireless communication system prototype}\label{fig:sys}
\end{figure}
As shown in Fig. \ref{fig:sys}, an end-to-end wireless communication system prototype is first established to collect the dataset,
	which consists of a source, a RF vector signal
	generator, a transmitter antenna, a receiver antenna, and a vector signal analyzer.
	The parameters of the devices of the proposed end-to-end wireless communication system prototype are listed in Table \ref{tab:type}.
	
	\begin{table}[h]
		\footnotesize
		\centering
		\caption{Experimental equipment and parameters}
		\label{tab:type}
		\begin{tabular}{l|lll}
			\hline\\[-2mm]
			Experiment setup & Type and parameters  \\
			\hline\\[-2mm]
			EXG RF vector signal generator &  Keysight N5172B  \\[1mm]
			MXA vector signal analyzer & Keysight N9020B  \\[1mm]
			Antenna Gain  & 24 ${{\rm{dBi}}}$  \\[1mm]
			\hline
		\end{tabular}
	\end{table}
	
	The volume environment is a $15 \times 5 \times3$ $\left( {{{\rm{m}}^3}} \right)$ office, where $15$, $5$, and $3$ denote the length, width, and height, respectively. 
	Note that the distance between the transmitter and the receiver is approximately $10$ meters. The power of the background noise is $78$ dBm.
	
	The carrier frequency ${f_c}$ and the sampling rate ${f_s}$ are $2.4$ GHz and $100$ MHz/s, respectively.
	For each $M$-QAM modulation scheme, the number of sample points $N$ has four cases, i.e., $N = {10, 20, 40}$, and $80$.
	
	To reduce the generalization error, the collected data set contains $10000$ transmit signal periods, in which $8000$ periods are used for training and $2000$ periods are used for testing.
	



	

	\subsection{Experimental Results}
	
	DBN-SVM based and AdaBoost based demodulators are trained on these training sets. The DBN-SVM based demodulator training ends after $110$ epochs, after which the training loss almost does not decline, and the AdaBoost based demodulator training ends when the iteration error is less than $10^{-3}$.
	In the experiment, signal sets with different SNRs, ranging from $3$ to $25$ dB, are chosen as the validation sets; the DBN based, SVM based, and MLD based demodulation methods are used for comparison.
	
	In Fig. \ref{fig 4qam} and Fig. \ref{fig 16qam}, the demodulation performance versus SNR of the proposed demodulator and the three baseline schemes are compared by the demodulation of $4$-QAM and $16$-QAM, respectively. The demodulation accuracy of the models increases as SNR increases.
	In particular, Fig. \ref{fig 4qam} indicates that the demodulation accuracy of all methods are close to $100\%$ when SNR $\ge 15$dB, and the proposed AdaBoost based demodulator is significantly superior to  the other models when SNR $\le13$dB. Besides, the proposed DBN-SVM based demodulator has better performance than the DBN-based and SVM-based demodulation methods.
	In Fig. \ref{fig 16qam}, compared with Fig. \ref{fig 4qam}, we focus on the same performance index at $16$-QAM. It shows the designed AdaBoost based demodulator is close to $100\%$ when SNR $\ge15$dB. However, other methods cannot approach $100\%$ as SNR increases. Furthermore, among these demodulation methods, the AdaBoost based demodulator obviously outperform the other four methods.
	It can be observed that the demodulation accuracy achieved by DBN-SVM based demodulator exceeds ones by the DBN-based, SVM-based demodulation methods. Although the overall trend of MLD classification accuracy increases as SNR increases, it has a obvious fluctuation. The reason is that the practical wireless channels include complicaful interferences, but the robustness of MLD is poor.
	
	
	{\color {blue}
	\begin{figure}[h]
		\centering
		\includegraphics[width=0.43\textwidth]{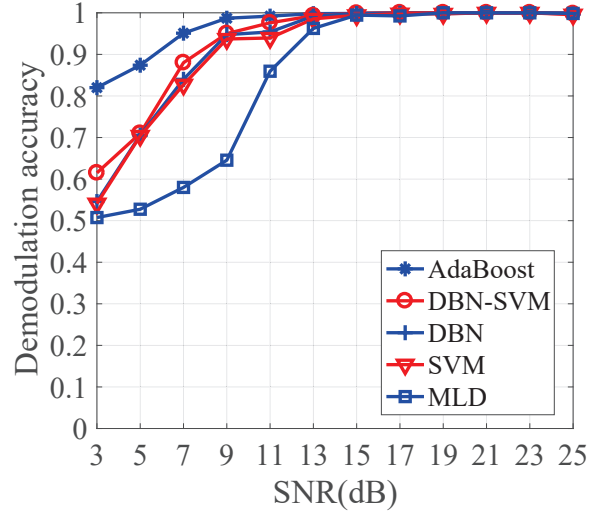}
		\caption{Demodulation accuracy comparison of different demodulators with $N = 40$ and  $4$-QAM}
		\label{fig 4qam}
	\end{figure}

	\begin{figure}[tbp]
		\centering
		\includegraphics[width=0.43\textwidth]{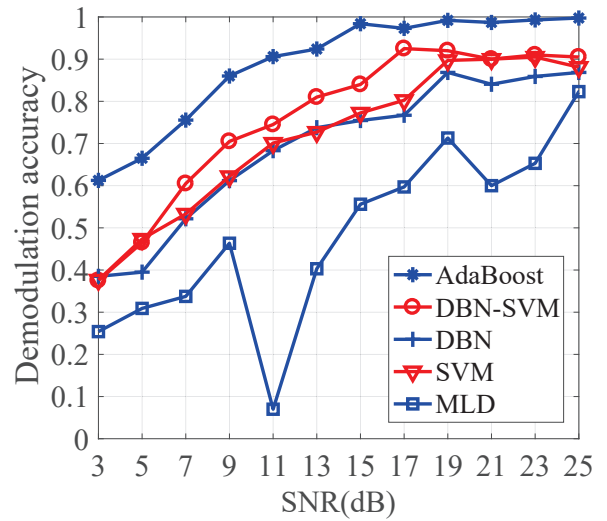}
		\caption{Demodulation accuracy comparison of different demodulators with $N = 40$ and  $16$-QAM}
		\label{fig 16qam}
	\end{figure}
}
	In Fig. \ref{diffsample}, the accuracy performance for different sampling points at $16$-QAM is simulated. It can be observed that the demodulation accuracy increases with the number of sample points. Furthermore, the demodulation accuracy can approximately achieve $100\%$ with $N=40$ or $N=80$ when SNR $\ge15$dB. However, with an increase in the number of sample points, the computational complexity also increases.

	
	\begin{figure}[h]
		\centering
		\includegraphics[width=0.44\textwidth]{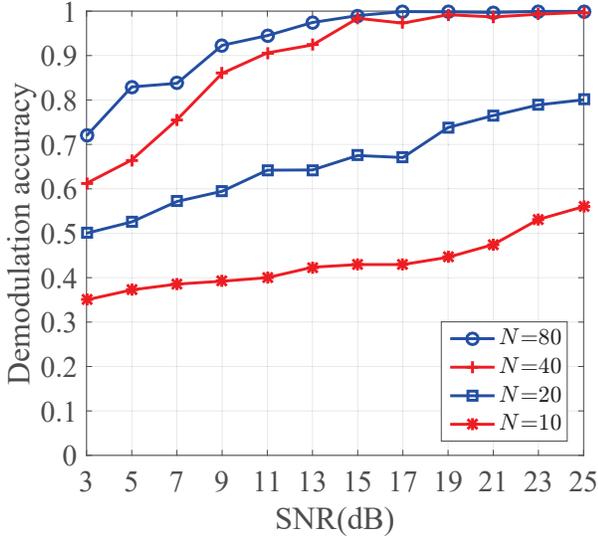}
		\caption{Demodulation accuracy of AdaBoost versus SNR with $16$-QAM}
		\label{diffsample}
	\end{figure}

\begin{figure}[tbp]
		\centering
		\includegraphics[width=0.43\textwidth]{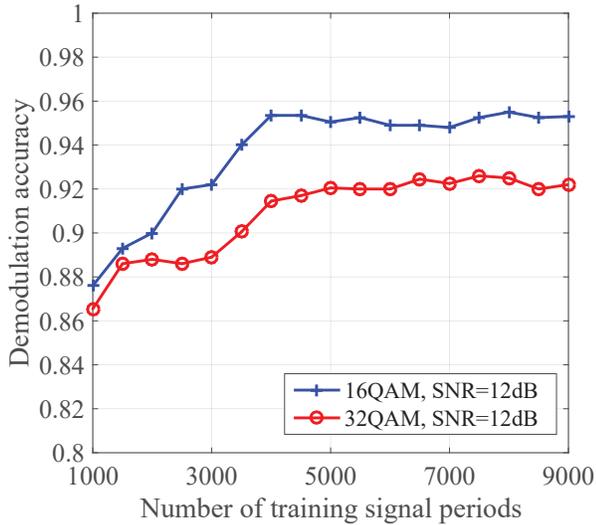}
		\caption{Demodulation performance versus different number of training signal periods with $16$-QAM and $32$-QAM}
		\label{diff_len_1632}
	\end{figure}

	Fig. \ref{diff_len_1632} shows the demodulation accuracy achieved by the AdaBoost based demodulator versus the number of training signal periods, where the number of sampling points is $40$ and SNR = $12$dB.
	The result shows that the demodulation accuracy initially increases with an increase in the number of training signal periods,
	and then, it reaches saturation when the number of training signal periods is $5000$. It can be observed that, compared with $32$-QAM, $16$-QAM can achieve higher accuracy. Meanwhile, $16$-QAM can provide stable performance with relatively fewer training signal periods.
	Different modulation models have different requirements with different number of training signals periods. In general, higher orders require longer training signals periods.
	
	
\begin{figure}[tbp]
		\centering
		\includegraphics[width=0.43\textwidth]{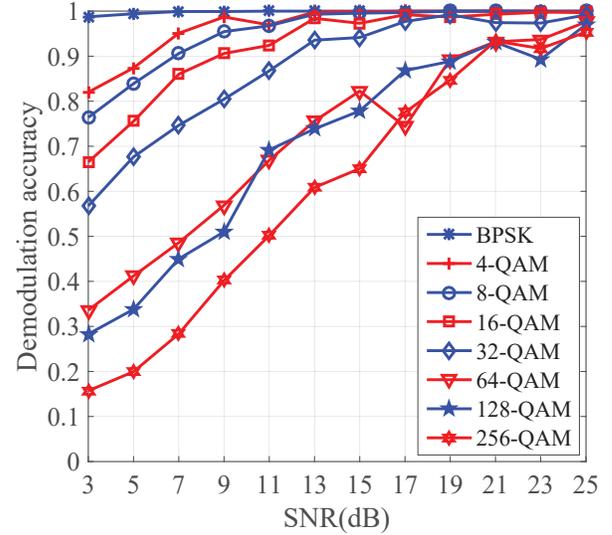}
		\caption{Demodulation accuracy comparison of different modulation mode versus SNR with AdaBoost based demodulator}
		\label{diff_modem}
	\end{figure}

	Fig. \ref{diff_modem} presents the demodulation accuracies of BPSK and $M$-QAM modulation schemes. In this experiment, the AdaBoost based demodulation algorithm was employed, where the number of sampling points is $N=40$. The demodulation accuracy for all modulation schemes increases with SNR. Meanwhile, the accuracy achieved by the BPSK-modulation scheme is better than the other seven schemes for the same SNR. Furthermore, Fig. \ref{diff_modem} also indicates that the demodulation accuracy reduces with an increase of the modulation order.
	
\begin{figure}[tbp]
		\centering
		\includegraphics[width=0.43\textwidth]{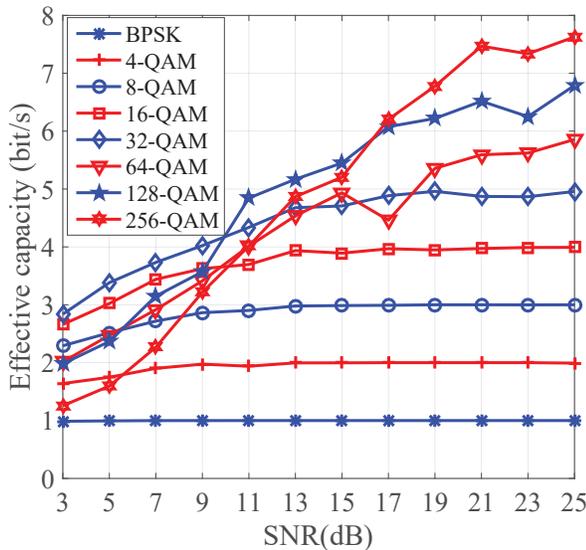}
		\caption{Effective capacity comparison of different modulation mode versus SNR with AdaBoost based demodulator}
		\label{diff_modem_ec}
	\end{figure}

	In Fig. \ref{diff_modem_ec}, the same modulation schemes, demodulation algorithm, and sampling points are used as in Fig. \ref{diff_modem}, where  the effective capacity of different modulation methods versus SNR are reported. The effective capacity by 
	 BPSK, $4$-QAM, and $8$-QAM almost remain unchanged with an increase in SNR. It is found that the modulation order has a considerable positive impact on the performance of the transmission capacity. The performance gap between the low order and the high order modulation is clearer when SNR $\leq 15$dB. However, the demodulation accuracy of high order modulation is low, so there is a trade-off between the demodulation accuracy and the effective capacity.
	
	
	\section{Conclusion}
	
	In this paper, a flexible end-to-end wireless communications prototype platform was proposed for real physical environments.   Then,  the first open measured modulation data dataset with eight modulation schemes, i.e., BPSK, $4$-QAM, $8$-QAM, $16$-QAM, $32$-QAM, $64$-QAM, $128$-QAM, and $256$-QAM, were established and accessed online. Furthermore, two DL-based demodulators, i.e., DBN-SVM based demodulator and AdaBoost based demodulator, were proposed. Based on the real dataset, the demodulation performance of the proposed demodulators were tested. Finally, experimental results indicated that the proposed demodulators outperform the DBN based, SVM based, and MLD based demodulators at various scenarios.

	\bibliographystyle{IEEE-unsorted}
	\bibliography{WIFI_0220}

\end{document}